\documentclass[10pt,fleqn]{article}

\usepackage{amssymb}

\newcommand{\name}[1]{\begin{flushleft}
                       \LARGE \bf #1
                       \end{flushleft}\vspace{-3mm}}

\newcommand{\Author}[1]{\begin{flushleft}
                       \it #1 \end{flushleft}}

\newcommand{\Adress}[1]{\begin{flushleft}
                       \it #1 \end{flushleft}}

\newcommand{\AbsEng}[1]{
    \begin{flushright}
    \begin{minipage}{120mm}
     \small   #1
    \end{minipage}
    \end{flushright}
}

\newcommand{\be}{\begin{equation}}
\newcommand{\ee}{\end{equation}}
\newcommand{\ba}{\hspace*{-5pt}\begin{array}}
\newcommand{\ea}{\end{array}}
\newcommand{\p}{\partial}
\newcommand{\ds}{\displaystyle}

\newcommand{\pbf}[1]{\mbox{\mathversion{bold}$#1$}}

\begin{document}

\name{On two-component equations \\
for zero mass particles}

\medskip

\noindent{published as Preprint of Institute of Theor. Phys., N~70-88E,
Kyiv, 1970, 22~p.}

\Author{Wilhelm I. FUSHCHYCH and A.L. GRISHCHENKO}

\Adress{Institute of Mathematics of the National Academy of
Sciences of Ukraine, \\ 3 Tereshchenkivska  Street, 01601 Kyiv-4,
UKRAINE}

\noindent {\tt URL:
http://www.imath.kiev.ua/\~{}appmath/wif.html\\ E-mail:
symmetry@imath.kiev.ua}

\AbsEng{The paper presents a detailed theoretical-group analysis of three types of
two-component equations of motion which describe the particle with zero mass
and spin~$\frac 12$. There are studied $P$-,  $T$- and $C$-propertias of the equations obtained.}

\medskip

\renewcommand{\theequation}{2.\arabic{equation}}
\setcounter{equation}{0}

\centerline{\bf 1. Introduction}

In the previous paper [1] it was shown by one of the authors that starting from
the four-component Dirac equation with zero mass one can obtain three types of
two-component equations. One of them coincides with the Weyl equation which, as is known, is
$P^{(k)}C$- and $T^{(1)}$-invariant but $P^{(k)}$-, $C$-noninvariant. Two other equations are
noninvariant with respect to $P^{(k)}C$-transformations. For one of these two equa\-tions the
$PTC$ theorem is not valid, i.e. such an equation is noninvariant with respect to
$P^{(k)}T^{(1)}C$- and $P^{(k)}T^{(2)}C$-transformations\footnote{Notations and definitions
which are gives without explations are the same as in the paper [1].}.

This present paper is dedicated to the detailed study of all possible (with an accuracy of
the unitary equivalence) two-component and four-component (with sub\-si\-dia\-ry conditions)
equations describing free notion of a particle with zero mass and spin $s=\frac 12$.

From the point of view of ideology the previous and the present papers are closely connected
with the papers by Shirokov~[2] and Foldy~[3] in which for the first time equations of motion
for a particle without antiparticle with non-zero mass and arbitrary spin were suggested.
The Shirokov--Foldy equations are $P^{(k)}$- and $T^{(1)}$-invariant, but $T^{(2)}$-
and $C$-noninvariant.

\medskip

\centerline{\bf 2. Three types of two-component equations}

{\bf 1.} The helicity and energy sign [2] operators~[2]
\be
\Lambda =\frac{J_{12}P_3 +J_{23} P_1 +J_{32}P_1}{E},
\qquad E=\sqrt{p_1^2 +p_2^2 +p_3^2}, \qquad \hat \varepsilon=\frac{P_0}{E}
\ee
are the Casimir operators of the group $P(1,3)$ for the representations with zero mass and
discrete spin.

Between the operators $P$, $T$, $C$ and $\Lambda$, $\hat \varepsilon$
it is easy to establish such relations\footnote{The results of this subsection are valid
for the arbitrary spin.}:
\be
P^{(k)} \Lambda =-\Lambda P^{(k)}, \qquad
P^{(k)} \hat \varepsilon =\hat \varepsilon P^{(k)}, \qquad k=1,2,3,
\ee
\be
T^{(a)} \Lambda =\Lambda T^{(a)}, \quad a=1,2, \qquad T^{(1)}\hat \varepsilon=
\hat \varepsilon T^{(1)}, \qquad
T^{(2)}\hat \varepsilon= -\hat \varepsilon T^{(2)},
\ee
\be
C\Lambda = \Lambda C, \qquad C \hat \varepsilon =-\hat \varepsilon C.
\ee

\newpage

Hence it follows such coupling scheme of irreducible representations of the proper
Poincar\'e group by the operators $P$, $T$, $C$:

\vspace*{-6mm}

\begin{center}
\unitlength 1.00mm
\linethickness{0.4pt}
\begin{picture}(80,30)
\put(35,21){$T^{(2)}P^{(k)}$}
\put(32,15){$C,T^{(2)}$}
\put(32,7){$C,T^{(2)}$}
\put(40,20){\vector(1,0){15}}
\put(40,20){\vector(-1,0){15}}
\put(28,0){$T^{(2)}P^{(k)}, CP^{(k)}$}
\put(40,4){\vector(1,0){15}}
\put(40,4){\vector(-1,0){15}}
\put(15,19){$D^+(s)$}
\put(56,19){$D^+(s)$}
\put(18,12){\vector(0,1){5}}
\put(18,12){\vector(0,-1){5}}
\put(12,3){$D^+(-s)$}
\put(56,3){$D^-(s)$}
\put(60,12){\vector(0,1){5}}
\put(60,12){\vector(0,-1){5}}
\put(40,12){\vector(3,1){15}}
\put(40,12){\vector(-3,-1){15}}
\put(40,12){\vector(3,-1){15}}
\put(40,12){\vector(-3,1){15}}
\put(11,11){$P^{(k)}$}
\put(61,11){$P^{(k)}$}
\end{picture}
\end{center}

It is seen from the scheme (2.5) that there exist three essentially different (with respect to
$P$-, $T$ and $C$-transformations) types of two-component equations of motion on the
solutions of which the following representations of the $P(1,3)$
group are realized:
\setcounter{equation}{5}
\be
D^+(s)\oplus D^- (-s) \qquad \mbox{or} \qquad D^-(s) \oplus D^+(-s),
\ee
\be
D^+(s)\oplus D^- (s)\phantom{-} \qquad \mbox{or} \qquad D^-(-s) \oplus D^+(-s),
\ee
\be
D^+(s)\oplus D^+ (-s) \qquad \mbox{or} \qquad D^-(s) \oplus D^-(-s).
\ee
Hence it follows such result:

\vspace{-2.5mm}

\begin{enumerate}
\item[1)]  the space $R_1$ where the representation (2.6) is realized is invariant with respect
to $T^{(1)}$- and $CP^{(k)}$-transformations but noninvariant with respect to $T^{(2)}$-,
$P^{(k)}$- and $C$-transformations;

\vspace{-2.5mm}

\item[2)] the space $R_2$ where the representation (2.7) is realized is invariant with respect to
$T^{(1)}$-, $T^{(2)}$- and $C$-transformations but noninvariant with respect to $P^{(k)}$-
and $CP^{(k)}$-transformations;

\vspace{-2.5mm}

\item[3)] the space $R_3$ where the representation (2.8) is realized, is invariant with respect to
$P^{(k)}$- and $T^{(1)}$-transformations but noninvariant with respect to $T^{(2)}$-
and $C$-transformations.

\vspace{-2.5mm}

\end{enumerate}

The two-component equations the wave functions of which are transformed ac\-cor\-ding to the
representations (2.6)--(2.8), have the same $P$-, $T$-  and $C$-properties as the spaces
$R_1$, $R_2$, $R_3$ have.

{\bf 2.} The Dirac equation
\be
\gamma_\mu p^\mu \Psi(t,\vec x)=0, \qquad \mu=0,1,2,3
\ee
is transformed to the form
\be
i\frac{\p \Phi (t,\vec x)}{\p t} =\gamma_0 \Phi(t,\vec x),
\ee
\be
\Phi(t,\vec x) =U \Psi(t,\vec x)
\ee
with the help of unitary transformation [4]
\be
U=\frac{1}{\sqrt{2}} \left(1+\frac{\gamma_k p_k}{E}\right).
\ee

In the representation (1.2.6) for the Dirac matrices\footnote{See (2.6) in [1]}
where
\[
\gamma_0 =\left( \begin{array}{cc} \sigma_3 & 0\\ 0 & -\sigma_3 \end{array}\right)
\]
eq. (2.10) decomposes into two two-component system
\be
i\frac{\p \Phi_\pm(t,\vec x)}{\p t} =\pm \sigma_3 E\Phi_\pm(t,\vec x),
\ee
$\Phi_\pm(t,\vec x)$ are two-component wave functions.

If tor the Dirac matrices we choose the representation, where
\[
\gamma_0 =\left( \begin{array}{cc} 1 & 0\\ 0 & -1 \end{array}\right)
\]
then (2.10) decomposes into the system
\be
i\frac{\p \widetilde \Phi_\pm(t,\vec x)}{\p t} =\pm  E\widetilde \Phi_\pm(t,\vec x),
\ee
$\widetilde \Phi_\pm(t,\vec x)$ are two-component wave functions.

Eqs. (2.13), (2.14) in themselves (without algebra $P(1,3)$)
do not unambiguously determine what particle and antiparticle they describe. Depending on
the rep\-re\-sen\-ta\-tion of the group $P(1,3)$ with respect to which its wave function is
transformed
under transformations from the group $P(1,3)$, the same (by the form) two-component equation
of motion describes, as is seen below, different particles. In other words, it means that the
equations of motion only together with the algebra $P(1,3)$ un\-am\-bi\-gu\-ous\-ly determine what
particle is described by it.

According to the results of the previous subsection for the particle with spin \mbox{$s=\frac 12$}
there are three essentially various two-dimension representations for the algebra $P(1,3)$.
They have the following form
\be
\ba{l}
\ds P_0^{\Phi_1} ={\mathcal H}^{\Phi_1} =\sigma_3 E, \qquad P_k^{\Phi_1}=p_k=-i\frac{\p}{\p x_k},
\vspace{2mm}\\
\ds J_{12}^{\Phi_1}=M_{12}+\frac{e_3 {\mathcal H}^{\Phi_1}}{2E}, \qquad
e_3 =\frac{p_3}{|p_3|}, \qquad p_3\not= 0,
\vspace{2mm}\\
\ds  J_{13}^{\Phi_1}=M_{13}-\frac{p_2 {\mathcal H}^{\Phi_1}}{2E(E+|p_3|)}, \qquad
J_{23}^{\Phi_1}=M_{23}+\frac{p_1 {\mathcal H}^{\Phi_1}}{2E(E+|p_3|)},
\vspace{2mm}\\
\ds J_{01}^{\Phi_1}=t_0 p_1 -\frac 12 \left[x_1,{\mathcal H}^{\Phi_1}\right]_+
-\frac{p_2 e_3}{2(E+|p_3|)},
\vspace{2mm}\\
\ds J_{02}^{\Phi_1}=t_0 p_2 -\frac 12 \left[x_2,{\mathcal H}^{\Phi_1}\right]_+
+\frac{p_1 e_3}{2(E+|p_3|)},
\vspace{2mm}\\
\ds J_{03}^{\Phi_1}=t_0 p_3 -\frac 12 \left[x_3,{\mathcal H}^{\Phi_1}\right]_+;
\ea
\ee
\be
\ba{l}
\ds P_0^{\Phi_2} ={\mathcal H}^{\Phi_2} =\sigma_3 E, \qquad P_k^{\Phi_2}=p_k,
\qquad
\ds J_{12}^{\Phi_2}=M_{12}+\frac{e_3}{2E},
\vspace{2mm}\\
\ds  J_{13}^{\Phi_2}=M_{13}-\frac{p_2}{2(E+|p_3|)}, \qquad
J_{23}^{\Phi_2}=M_{23}+\frac{p_1 }{2(E+|p_3|)},
\vspace{2mm}\\
\ds J_{01}^{\Phi_2}=t_0 p_1 -\frac 12 \left[x_1,{\mathcal H}^{\Phi_2}\right]_+
-\frac{p_2 e_3{\mathcal H}^{\Phi_2}}{2E(E+|p_3|)},
\vspace{2mm}\\
\ds J_{02}^{\Phi_2}=t_0 p_2 -\frac 12 \left[x_2,{\mathcal H}^{\Phi_2}\right]_+
+\frac{p_1 e_3{\mathcal H}^{\Phi_2}}{2E(E+|p_3|)},
\vspace{2mm}\\
\ds J_{03}^{\Phi_2}=t_0 p_3 -\frac 12 \left[x_3,{\mathcal H}^{\Phi_2}\right]_+;
\ea
\ee
\be
\ba{l}
\ds P_0^{\Phi_3} =E={\mathcal H}^{\Phi_3} \qquad P_k^{\Phi_3}=p_k,
\qquad
 J_{12}^{\Phi_3}=M_{12}+\frac{e_3\sigma_3}{2},
\vspace{2mm}\\
\ds  J_{13}^{\Phi_3}=M_{13}-\frac{p_2\sigma_3}{2(E+|p_3|)}, \qquad
J_{23}^{\Phi_3}=M_{23}+\frac{p_1 \sigma_3}{2(E+|p_3|)},
\vspace{2mm}\\
\ds J_{01}^{\Phi_3}=t_0 p_1 -\frac 12 \left[x_1,{\mathcal H}^{\Phi_3}\right]_+
-\frac{p_2 e_3\sigma_3}{2(E+|p_3|)},
\vspace{2mm}\\
\ds J_{02}^{\Phi_3}=t_0 p_2 -\frac 12 \left[x_2,{\mathcal H}^{\Phi_3}\right]_+
+\frac{p_1 e_3\sigma_3}{2(E+|p_3|)},
\vspace{2mm}\\
\ds J_{03}^{\Phi_3}=t_0 p_3 -\frac 12 \left[x_3,{\mathcal H}^{\Phi_3}\right]_+,
\ea
\ee
where $M_{kl}=x_k p_l -x_l p_k$.

By direct verification one can be convinced that the operators (2.15)--(2.17) satisfy the
commutation relations of algebra $P(1,3)$. These three representations are not equivalent.
Really the operators of energy sign and helicity have the form
\[
\ba{lll}
\ds \hat \varepsilon =\frac{{\mathcal H}^c}{E}=\sigma_3, &
\ds \Lambda =\frac12 \hat \varepsilon & \mbox{for the representation (2.15),}
\vspace{2mm}\\
\ds \hat \varepsilon =\sigma_3, & \ds \Lambda =\frac12 &
 \mbox{for the representation (2.16),}
\vspace{2mm}\\
\hat \varepsilon =1, & \ds  \Lambda =\frac12 \sigma_3 & \mbox{for the representation (2.17).}
\ea
\]
Hence it is clear that the representations (2.12), (2.16), (2.17) are not equivalent and are given
in the spaces $R_1$, $R_2$, $R_3$ respectively.

Besides two-dimensional representations given for the algebra $P(1,3)$ one can, evidently,
obtain the other ones as well which however, will be unitary-equivalent to (2.15)--(2.17).
If, for example, in (2.15)--(2.17) one performs the substitution
\be
e_3\to 1, \qquad |p_3|\to p_3,
\ee
then the operators obtained also realize the representations of the algebra $P(1,3)$.
The explicit form for the generators of the group $P(1,3)$ obtained from (2.15)--(2.17) with the
help of substitution~(2.18) will be denoted in the sequal by~($2.15'$)--($2.17'$).

If in (2.15)--(2.17) the matrix $\sigma_3$ is substituted by 1 (or $-1$),
then such operators will realize one-dimensional irreducible representations of the algebra
$P(1,3)$ which are, of course, unitarily equivalent to the corresponding one-dimensional
Shirokov representations~[5]. The representations~[5] are obtained without connection with
the equations of motion and are realized on the functions $\Psi(p_1,p_2,p_3)$
not depending on the time.

Summing up all the above presented we come to the conclusion:

\vspace{-2mm}

\begin{enumerate}
\item[1)] Eq. (2.13) together with algebra (2.15) (or ($2.15'$)) describes the particle with
helicity $+\frac 12$ and the antiparticle with helicity
$-\frac12$\footnote{The representation $D^+(s)$ corresponds to the particle and the
representation $D^-(s)$ to the antiparticle.};

\vspace{-2mm}

\item[2)] Eq. (2.13) together with algebra (2.16) (or ($2.16'$)) describes the particle with helicity
$+\frac 12$ and the antiparticle with helicity $+\frac 12$;

\vspace{-2mm}

\item[3)] Eq. (2.14) together with algebra (2.17) (or ($2.17'$)) describes two particles with helicity
$+\frac 12$ and $-\frac 12$\footnote{Eq. (2.14) can be interpreted as the equation of motion for
one particle which can be in two states differing one from another by the helicity sign.}.

\vspace{-2mm}

\end{enumerate}

If Eq. (2.13) is connected with the algebra (2.13) (or ($2.15'$)) the wave function of such
equation is denoted by $\Phi_1$ (or $\Phi_1'$). The wave function in Eq.~(2.13) connected with
the algebra~(2.16) (or~($2.16'$)) is denoted by $\Phi_2$ (or $\Phi_2'$). Similarly
$\Phi_3$ (or $\Phi_3'$) denotes the wave function in Eq.~(2.14) connected with the
algebra~(2.17) (or~($2.17'$)).

{\bf 3.} The transition from the canonical equation (2.13) to the non-canonical one of the
type~(1.3.1) is realized with the help of unitary transformation~[1]
\be
v_1^{-1} =\frac{E+|p_3|+i(\sigma_1 p_2 -\sigma_2 p_1)}{\{2E(E+|p_3|)\}^{1/2}}.
\ee
Under this transformation Eq. (2.13) takes the form
\be
i \frac{\p \chi(t,\vec x)}{\p t} =(\sigma_1 p_1 +\sigma_2 p_2 +\sigma_3 |p_3|)
\chi(t,\vec x),
\ee
where
\be
\chi=\chi_1 =v_1^{-1}\Phi_1
\ee
or
\be
\chi=\chi_2 =v_1^{-1}\Phi_2.
\ee

The type of Eq. (2.14) under transformation (2.19) is not changed. The operators (2.15), (2.16)
in $\chi$-representation have the form
\be
P_0^{\chi_a} ={\mathcal H}=\sigma_1 p_1+\sigma_2 p_2 +
\sigma_3 |p_3|, \qquad P_k^{\chi_a}=p_k , \qquad a=1,2,
\ee
\be
J_{\mu\nu}^{\chi_1} =J_{\mu\nu}^{\Phi_1}\left( x_k \to X_k, {\mathcal H}^{\Phi_1}\to
{\mathcal H}\right),
\ee
\be
J_{\mu\nu}^{\chi_2} =J_{\mu\nu}^{\Phi_2}\left( x_k \to X_k, {\mathcal H}^{\Phi_2}\to
{\mathcal H}\right),
\ee
where
\be
\ba{l}
\ds X_1=x_1 -\frac{\sigma_2}{2E} +\frac{\sigma_3 p_2}{2E(E+|p_3|)}-
\frac{p_1(\sigma_1p_2 -\sigma_2 p_1)}{2E^2(E+|p_3|)},
\vspace{2mm}\\
\ds X_2=x_2 +\frac{\sigma_1}{2E} -\frac{\sigma_3 p_1}{2E(E+|p_3|)}-
\frac{p_2(\sigma_1p_2 -\sigma_2 p_1)}{2E^2(E+|p_3|)},
\vspace{2mm}\\
\ds X_3=x_3- \frac{\sigma_1p_2 -\sigma_2 p_1}{2E^2}e_3.
\ea
\ee

If one connects with Eqs. (2.13) and (2.14) the representations ($2.15'$)--($2.17'$),
but not the representations (2.l5)--(2.17) in this case the transition from the canonical
equation to the noncanonical one can he conveniently realized with the help of the unitary
transformation~[6]
\be
v^{-1} =\frac{E+p_3 +i(\sigma_1 p_2 -\sigma_2 p_1)}{\{2E(E+|p_3|)\}^{1/2}}.
\ee

Under this transformation Eq. (2.13) takes the form of the Weyl equation
\be
i\frac{\p \chi^w(t,\vec x)}{\p t} =(\sigma_1 p_1 +\sigma_2 p_2 +\sigma_3 p_3)\chi^w(t,\vec x),
\ee
where
\be
\chi^w\equiv \chi_1^w =v^{-1}\Phi_1'
\ee
or
\be
\chi^w\equiv \chi_2^w =v^{-1}\Phi_2'.
\ee
Eq. (2.14) is unchanged by the transformation (2.27). The generators
$J_{\mu\nu}^{w_2}\equiv J_{\mu\nu}^{\chi_2^w}$ coincide with~(2.25) where the operator
${\mathcal H}$ has the form~(2.31) and in the operators (2.26) the substitution~(2.18)
is performed. This algebra is denoted by ($2.25'$). Under transformation~(2.27) the
algebra~($2.15'$) goes into the algebra
\be
\ba{l}
P_0^{w_1} ={\mathcal H}=\sigma_k p_k, \qquad P_k^{\chi_1^w}=p_k,
\vspace{2mm}\\
\ds J_{kl}^{w_1}=M_{kl}+ \frac{\sigma_n}{2}, \qquad k,l,n \quad \mbox{is the cycle (1,2,3)},
\vspace{2mm}\\
\ds J_{0k}^{w_2} =t_0 p_k -\frac 12 [x_k,{\mathcal H}]_+.
\ea
\ee
The algebra ($2.17'$) is transformed into the algebra
\be
\ba{l}
P_0^{w_3} ={\mathcal H}=E, \qquad P_k^{w_3}=p_k,
\vspace{2mm}\\
\ds J_{kl}^{w_3}=M_{kl}+ \frac{\sigma_n}{2}, \qquad k,l,n \quad \mbox{is the cycle (1,2,3)},
\vspace{2mm}\\
\ds J_{0k}^{w_3} =t_0 p_k -\frac 12 [x_k,{\mathcal H}]_+ +\frac{\sigma_n p_l-\sigma_l p_n}{E}.
\ea
\ee
The operators $P_\mu^{w_k}$, $J_{\mu\nu}^{w_k}$ are defined on the corresponding sets
$\{\chi_k^w\}$, $k=1,2,3$.

From the above given analysis it follows such a result:

\vspace{-3mm}

\begin{enumerate}

\item[1)] Eq. (2.20) together with the algebra (2.24) (or Eq. (2.28) together with the algebra~(2.31))
describes the particle with helicity $+\frac 12$ and the antiparticle with helicity $-\frac 12$;

\vspace{-2mm}

\item[2)] Eq. (2.20) together with the algebra (2.25) (or eq. (2.28)) together with the
algebra ($2.25'$)) describes the particle with helicity $+\frac 12$ and the antiparticle with helicity
$+\frac 12$.

\vspace{-3mm}

\end{enumerate}

Thus, the same (by the form) Eq.(2.20) (or the Weyl equation (2.28)) describes different
types of particles and antiparticles depending on the representation of group $P(1,3)$
with respect to which the wave functions $\chi$ (or $\chi^w$) are transformed under
transformations from the group $P(1,3)$.

Not only the equations invariant with respect to the group $P(1,3)$
have such a dual nature but also the equations invariant with respect to the inhomogeneous
de Sitter group $P(1,4)$. Within the frameworks of the $P(1,4)$ group the same (by the form)
equation of the Dirac type describes the various types of particles and antiparticles~[7].

\medskip

\renewcommand{\theequation}{3.\arabic{equation}}
\setcounter{equation}{0}

\centerline{\bf \S~3. $\pbf{P}$-, $\pbf{T}$- and $\pbf{C}$-properties of two-component equations}

In studying $P$-, $T$- and $C$-properties of the equations of motion one does not indicate
as a rule, with what algebra $P(1,3)$ the given equation is connected. Such an approach, as
it follows from the results of the previous section, is not quite correct for the studying $P$-,
$T$, $C$-properties of Eqs.~(2.13), (2.20), (2.28), since the same equation connected with
various algebras $P(1,3)$ can have various properties with respect to the space-time reflections.

In order the equation, invariant with respect to the proper group $P(1,3)$, be $P$-, $T$-
and $C$-invariant it is necessary and sufficient to satisfy such relations:
\be
\ba{l}
\left[ P^{(k)},{\mathcal H}\right]_-=0, \qquad k=1,2,3,
\vspace{1mm}\\
\left[P^{(k)}, P_l\right]_-=0 \qquad \mbox{for} \quad k\not= l,
\vspace{1mm}\\
\left[P^{(k)}, P_l\right]_+=0 \qquad \mbox{for} \quad k= l,
\vspace{1mm}\\
\left[P^{(k)}, J_{lr}\right]_+=0 \qquad \mbox{for} \quad k= l, \ k=r,
\vspace{1mm}\\
\left[P^{(k)}, J_{lr}\right]_-=0 \qquad \mbox{for} \quad k\not= l, \ k\not=r,
\vspace{1mm}\\
\left[P^{(k)}, J_{0l}\right]_-=0 \qquad \mbox{for} \quad k\not= l,
\vspace{1mm}\\
\left[P^{(k)}, J_{0l}\right]_+=0 \qquad \mbox{for} \quad k= l;
\ea
\ee
\be
\left[ T^{(1)},{\mathcal H}\right]_-=\left[T^{(1)}, J_{0l}\right]_-=0,
\qquad \left[T^{(1)}, P_k\right]_+=\left[T^{(1)}, J_{kl}\right]_+=0,
\ee
\be
\left[ T^{(2)},{\mathcal H}\right]_+=\left[T^{(2)}, J_{0l}\right]_+=0,
\qquad \left[T^{(2)}, P_k\right]_-=\left[T^{(2)}, J_{kl}\right]_-=0,
\ee
\be
\left[C,{\mathcal H}\right]_+=\left[C, P_k\right]_+=\left[C, J_{\mu\nu}\right]_+=0.
\ee

Hence it follows that the equation of motion is invariant with respect to
$P$-trans\-for\-ma\-tion if all the conditions (3.1) are satisfied. Usually when studying
$P$-properties of the equations one verifies only the first relation from~(3.1) that, evidently,
is not sufficient for the correct conclusion.

How we give the explicit expressions for the operators $r^{(k)}$, $\tau^{(i)}$
(see formulas (1.3.2)--(1.3.5)) determining the operators of discrete transformations.

On the sets $\{\Phi_1\}$ and $\{\Phi'_1\}$ the operators $P^{(k)}$, $T^{(2)}$ and
$C$ cannot be determined since the range of values of these operators does not belong to
the sets $\{\Phi_1\}$ and $\{\Phi'_1\}$. The operator $T^{(1)}$ on
$\{\Phi_1\}$, $\{\Phi'_1\}$ can be defined and it is determined by such operators
\be
\tau^{(1)}=1 \qquad \mbox{or} \quad \sigma_3 \qquad \mbox{on} \quad \{\Phi_1\},
\ee
\be
\tau^{(1)}=\frac{\sigma_3 p_2 +ip_1}{\sqrt{p_1^2 +p_2^2}} \qquad
\mbox{on} \quad \{\Phi'_1\}.
\ee

$T^{(1)}$, $T^{(2)}$ and $C$ on the sets $\{\Phi_2\}$, $\{\Phi'_2\}$ are given by the operators
$\tau^{(k)}$, $k=1,2,3$
\be
\tau^{(2)}=\sigma_1 \qquad \mbox{or} \quad \sigma_2 \qquad \mbox{on} \quad \{\Phi_2\},
\ee
\be
\tau^{(3)}=\sigma_1 \qquad \mbox{or} \quad \sigma_2 \qquad \mbox{on} \quad \{\Phi_2\}.
\ee
The operator $\tau^{(1)}$ on the sets $\{\Phi_1\}$ and $\{\Phi_2\}$ has the form (3.5)
\be
\tau^{(1)}=\frac{p_2+ip_1}{\sqrt{p_1^2 +p_2^2}} \qquad \mbox{or} \quad
\frac{\sigma_3(p_2+ip_1)}{\sqrt{p_1^2 +p_2^2}} \qquad \mbox{on} \quad \{\Phi'_2\},
\ee
\be
\tau^{(2)}=\sigma_1 \qquad \mbox{or} \quad \sigma_2 \qquad \mbox{on} \quad \{\Phi'_2\},
\ee
\be
\tau^{(3)}=\sigma_1 \frac{p_2+ip_1}{\sqrt{p_1^2 +p_2^2}} \qquad \mbox{on} \quad
\{\Phi'_2\}.
\ee
The operators $P^{(k)}$ are not determined on $\{\Phi_2\}$ and $\{\Phi'_2\}$.
$P^{(k)}$ and $T^{(1)}$ on the sets $\{\Phi_3\}$, $\{\Phi'_3\}$ are given by:
\be
\tau^{(1)}=1 \qquad \mbox{or} \quad \sigma_3 \qquad \mbox{on} \quad \{\Phi_3\},
\ee
\be
r^{(k)}=\sigma_1 \qquad \mbox{or} \quad \sigma_2 \qquad \mbox{on} \quad \{\Phi_3\},
\ee
\be
\tau^{(1)}= \frac{p_2+i\sigma_3 p_1}{\sqrt{p_1^2 +p_2^2}} \qquad \mbox{on} \quad
\{\Phi'_3\},
\ee
\be
\ba{l}
r^{(1)}=\sigma_1 \qquad r^{(2)}= \sigma_2 \qquad \mbox{on} \quad \{\Phi'_3\},
\vspace{2mm}\\
\ds r^{(3)}= \frac{\sigma_a p_a}{\sqrt{p_1^2 +p_2^2}} \qquad \mbox{on} \quad
\{\Phi'_3\}.
\ea
\ee

The operators $r^{(k)}$ and $\tau^{(k)}$ on the sets $\{\chi\}$ and  $\{\chi^w\}$ have the form
\be
\ba{l}
\ds \tau^{(1)}=1-\frac{p_1^2}{E(E+|p_3|)}+\frac{ip_1p_2 \sigma_3}{E(E+|p_3|)}-
\frac{i\sigma_2 p_1}{E} \qquad \mbox{or}
\vspace{2mm}\\
\ds \tau^{(1)}=\sigma_3 \left( 1-\frac{p_2^2}{E(E+|p_3|)}\right) -\frac{ip_1p_2}{E(E+|p_3|)}+
\frac{\sigma_2 p_2}{E} \qquad \mbox{on} \quad \{\chi_1\}, \ \{\chi_2\},
\ea\hspace{-5pt}
\ee
\be
\tau^{(1)}=\sigma_2 \qquad \mbox{on} \quad \{\chi_1^w\},
\ee
\be
\ba{l}
\ds \tau^{(2)}= \left( 1-\frac{p_1^2}{E(E+|p_3|)}\right)\sigma_1 -\frac{\sigma_2p_1p_2}{E(E+|p_3|)}-
\frac{\sigma_3 p_1}{E} \qquad \mbox{or}
\vspace{2mm}\\
\ds \tau^{(2)}= \left( 1-\frac{p_2^2}{E(E+|p_3|)}\right)\sigma_2 -\frac{\sigma_1p_1p_2}{E(E+|p_3|)}-
\frac{\sigma_3 p_2}{E} \qquad \mbox{on} \quad \{\chi_2\},
\ea
\ee
\be
\tau^{(3)}=\sigma_1 \qquad \mbox{or} \qquad
\tau^{(3)}=\frac{\sigma_2|p_3|}{E} -\frac{p_2 \sigma_3 +ip_1}{E} \qquad
\mbox{on} \quad \{\chi_2\};
\ee
\be
\ba{l}
\ds \tau^{(1)} =\frac 1E (p_2-ip_1) (p_2^2 -i\sigma_2 p_1 E -i\sigma_1 p_2 p_3 +
i\sigma_1 p_1 p_2) \qquad \mbox{or}
\vspace{2mm}\\
\ds \tau^{(1)}=\frac 1E(-ip_1p_2 +\sigma_2 p_2 E-\sigma_1 p_1 p_3 +
\sigma_3 p_1^2)(p_2 -ip_1) \qquad \mbox{on}\quad \{\chi_2^w\};
\ea
\ee
\be
\ba{l}
\ds \tau^{(2)} =\sigma_1 -\frac{\sigma_3 p_1}{E}-\frac{p_1\sigma_a p_a}{E(E+|p_3|)}
\qquad \mbox{on} \quad \{\chi_2^w\}, \quad \mbox{or}
\vspace{2mm}\\
\ds \tau^{(2)} =\sigma_2 -\frac{\sigma_3 p_2}{E}-\frac{p_2\sigma_a p_a}{E(E+|p_3|)}
\qquad \mbox{on} \quad \{\chi_2^w\};
\ea
\ee
\be
\tau^{(3)}=\frac{p_2-ip_1}{E}\left\{ \sigma_1(p_1^2+p_2^2) -ip_2p_3 +
\sigma_3 p_1 p_3 \right\} \qquad \mbox{on} \quad \{\chi_2^w\};
\ee
\be
\tau^{(1)}=\frac 1E (p_2 -i\sigma_1 p_3 +i\sigma_3 p_1) \qquad \mbox{on} \quad
\{\chi_3^w\};
\ee
\be
r^{(k)}=\sigma_k, \qquad k=1,2,3 \qquad \mbox{on} \quad \{\chi_3^w\}.
\ee

The operators (3.16)--(3.24) are obtained from (3.5)--(3.11), (3.14),(3.17) with the help of
transformations (2.19), (2.27). The transformation law of these operators is given in (D.10)--(D.16).

Summing up all the above said we come to the final conclusion:

\vspace{-2mm}

\begin{enumerate}
\item[1)] Eq. (2.13) for the function $\Phi_1$ (or $\Phi'_1$) is $T^{(1)}$- and
$P^{(k)}C$-invariant, but $P^{(k)}$-, $T^{(2)}$- and $C$-noninvariant;

\vspace{-2mm}

\item[2)] Eq. (2.13) for the function $\Phi_2$ (or $\Phi'_2$) is $T^{(1)}$-,
$T^{(2)}$- and $C$-invariant, but $P^{(k)}$- and  $CP^{(k)}$-noninvariant;

\vspace{-2mm}

\item[3)] Eq. (2.14) is $P^{(k)}$- and $T^{(1)}$-invariant, but $T^{(2)}$-,  $C$- and
$CP^{(k)}$-noninvariant.

\vspace{-2mm}

\end{enumerate}

Evidently Eqs. (2.20), (2.28) have these properties as well.

\medskip

\noindent
{\bf Note 1.} In [1] we established $P$-, $T$- and $C$-properties of Eq.~(2.20) starting from the
assumption that $r^{(k)}$, $\tau^{(k)}$ on the set $\{\chi\}$ are the $2\times 2$ matrices. As is
seen from the previous such an assumption is limited. On the set $\{\chi\}$
$r^{(k)}$, $\tau^{(k)}$ are the operator functions depending on the momentum components
of the particle.

\medskip

\noindent
{\bf Note 2.} Under the four-dimensional rotations in Minkovski space the wave functions
$\Phi_1$, $\chi_1$, $\Phi_2$, $\chi_2^w$, $\Phi_3$, $\chi_3$, $\chi_2$ are transformed
nonlocally.

\medskip

In conclussion of this section we give some corrollaries immediately following from the
previous, which can be useful for the construction of weak interaction models on the basis
of the equations obtained.

\medskip

\noindent
{\bf Corrollary 1.} {\it  Any (one-component or two-component) equation of motion for the
particles with zero mass is Invariant with respect to the Wigner reflection of time $T^{(1)}$.}

\medskip

\noindent
{\bf Corrollary 2.} {\it  Eq. (2.20) for the function $\chi_2$ (or (2.28) for the function $\chi_2^w$)
is $T^{(1)}C$- and $T^{(2)}C$-invariant, but $PC$-,  $PT^{(1)}$-,
$PT^{(2)}$-, $PT^{(1)}C$- and $PT^{(2)}C$-noninvariant. It means that for such equation
neither hypothesis of combined parity conservation, nor hypothesis of $PTC$-invariance
conservation is valid.}

\medskip

\noindent
{\bf Corrollary 3.} {\it Eq. (2.14) is $PT^{(1)}$-, $T^{(2)}C$- and $PT^{(2)}C$-invariant, but
$PT^{(2)}$-, $PC$- and $PT^{(2)}C$-noninvariant.}

\bigskip

\renewcommand{\theequation}{4.\arabic{equation}}
\setcounter{equation}{0}

\centerline{\bf \S~4. $\pbf{CP}$-noninvariant subsidiary conditions}

The results of previous sections can be rather simply and briefly formulated if one describes
the zero mass particle with the help of four-component wave function. In this case the wave
function has the redundant (nonphysical) components which ran be invariantly separated
with the help of relativistic-invariant subsidiary conditions. From \S~2,~3
it follows that there are three types of subsidiary conditions. One of them is well known and
has the form
\be
{\mathcal P}_1^+ \Psi =0 \qquad \mbox{or} \qquad {\mathcal P}_1^- \Psi =0,
\ee
\be
{\mathcal P}_1^\pm =\frac 12 (1\pm \gamma_4).
\ee

Eq. (2.9) together with the condition (4.1) is equivalent to Eq. (2.28) for the function
$\chi_1^w$ (or (2.20) for the function $\chi_1$).

Now we find two other relativistic-invariant subsidiary conditions. Besides the matrix
$\gamma_4$ the energy sign operator commutes with the algebra (1.2.1). Hence it is clear
that the operators
\be
{\mathcal P}_2^\pm =\frac 12 \left( 1\pm \gamma_4 \frac{{\mathcal H}}{E}\right), \qquad
{\mathcal H}=\gamma_0 \gamma_k p_k,
\ee
\be
{\mathcal P}_3^\pm =\frac 12 \left( 1\pm \frac{{\mathcal H}}{E}\right)
\ee
commute with the algebra (1.2.1). The operators ${\mathcal P}_2^\pm$,
${\mathcal P}_3^\pm$ are the projection operators. We show then that the conditions
\be
{\mathcal P}_2^+ \Psi =0 \qquad \mbox{or} \qquad {\mathcal P}_2^- \Psi =0,
\ee
\be
{\mathcal P}_3^+ \Psi =0 \qquad \mbox{or} \qquad {\mathcal P}_3^- \Psi =0,
\ee
can be considered as subsidiary conditions.

Between the operators ${\mathcal P}_2^\pm$, ${\mathcal P}_3^\pm$ and $P$, $T$, $C$
it is easy to establish the following relations:
\be
\ba{ll}
P^{(k)}{\mathcal P}_2^\pm ={\mathcal P}^\mp_2 P^{(k)}, \qquad &
T^{(1)}{\mathcal P}_2^\pm ={\mathcal P}^\pm_2 T^{(1)},
\vspace{2mm}\\
T^{(2)}{\mathcal P}_2^\pm ={\mathcal P}^\pm_2 T^{(2)}, \qquad &
C{\mathcal P}_2^\pm ={\mathcal P}^\pm_2 C,
\ea
\ee
\be
\ba{ll}
P^{(k)}{\mathcal P}_3^\pm ={\mathcal P}^\pm_3 P^{(k)}, \qquad &
T^{(1)}{\mathcal P}_3^\pm ={\mathcal P}^\pm_3 T^{(1)},
\vspace{2mm}\\
T^{(2)}{\mathcal P}_3^\pm ={\mathcal P}^\mp_3 T^{(2)}, \qquad &
C{\mathcal P}_3^\pm ={\mathcal P}^\mp_3 C.
\ea
\ee

From (4.7), (4.8) it follows that the condition (4.5) is $T^{(1)}$-, $T^{(2)}$- and
$C$-invariant, but $P^{(k)}$- and $CP^{(k)}$-noninvariant, and the condition~(4.6) is
$P^{(k)}$- and $T^{(1)}$-invariant, but $T^{(2)}$- and $C$-noninvariant. It means that the
representation (2.7) is realized on the set ${\mathcal P}_2^\pm\{\Psi\}$,
and the representation~(2.8) is realized on the set ${\mathcal P}_3^\pm$.

Thus we came to the following result:

\vspace{-2mm}

\begin{enumerate}
\item[1)] Eq. (2.9) with subsidiary condition (4.5) is $T^{(1)}$-, $T^{(2)}$- and $C$-invariant,
but $P^{(k)}$-, $CP^{(k)}$-, $P^{(k)}T^{(1)}C$- and $P^{(k)}T^{(2)}C$-noninvariant;

\vspace{-2mm}

\item[2)] Eq. (2.9) with subsidiary condition (4.6) is $P^{(k)}$-, $T^{(1)}$- and
$P^{(k)}T^{(2)}C$-invariant, but $T^{(2)}$-, $C$-, $P^{(k)}T^{(1)}C$- and $P^{(k)}C$-noninvariant.

\vspace{-2mm}

\end{enumerate}

Eq. (2.9) with subsidiary conditions (4.1), (4.5), (4.6) can be written in the form of three equations
\be
\left(\gamma_\mu p^\mu +\varkappa_k {\mathcal P}_k^+\right){\mathcal P}_k^- \Psi(t,x)=0,
\qquad k=1,2,3,
\ee
where $\varkappa_k$, $k=1,2,3$ are the arbitrary constant numbers. For eqs. (4.9) the
conditions (4.1), (4.5), (4.6) are satisfied automatically.

\medskip

\renewcommand{\theequation}{{\rm D}.\arabic{equation}}
\setcounter{equation}{0}

\centerline{\bf Appendix}

In this appendix we present the main formulas according to which the operators
$r$, $r^{(k)}$, $\tau^{(k)}$ in representations $\{\Phi\}$ and $\{\chi\}$ were calculated
(see (3.5)--(3.25)).

To make it complete we give a definition to the combined parity
\be
CP\Phi (t,\vec x) =\theta \Phi^*(t,-\vec x), \qquad
CP^{(k)} \Phi(t,\vec x) =\theta^{(k)} \Phi^* (t,-x_k),
\ee
\be
P\Phi(t,\vec x)=r \Phi(t,-\vec x), \qquad P\equiv P^{(1)} P^{(2)} P^{(3)}.
\ee

From (3.1)--(3.4) and from the definitions (D.1), (D.2), (1.3.2)--(1.3.5) we obtain such relations
\be
\ba{l}
[r,p_k]_+=0, \qquad r{\mathcal H}(-\vec p)-{\mathcal H}(\vec p)r=0,
\vspace{1mm}\\
rJ_{kl}(-\vec x)-J_{kl}(\vec x) r=0, \qquad
rJ_{0k}(-\vec x)+J_{0k}(\vec x) r=0,
\ea
\ee
\be
\ba{l}
r(\vec p) r(-\vec p)=1, \qquad [r^{(k)}, p_n]_\pm=0, \qquad
r^{(k)} {\mathcal H}(-p_k) -{\mathcal H}(-p_k) r^{(k)}=0,
\vspace{1mm}\\
r^{(k)}J_{nl}(- x_k)\pm J_{nl}( x_k) r^{(k)}=0, \qquad
r^{(k)}J_{0n}(- x_k)\pm J_{0n}( x_k) r^{(k)}=0,
\ea
\ee
where ``$+$'' is taken if $k=n$ or $k=l$;
\be
\ba{l}
r^{(k)}(p_k) r^{(k)} (-p_k)=1, \qquad \tau^{(1)} {\mathcal H}^* -{\mathcal H} \tau^{(1)}=0,
\qquad \tau^{(1)} p_k^* +p_k \tau^{(1)}=0,
\vspace{1mm}\\
\tau^{(1)} J^*_{kl}+J_{kl} \tau^{(1)}=0, \qquad
\tau^{(1)} J^*_{0k}(-t_0)-J_{0k}(t_0) \tau^{(1)}=0,
\ea
\ee
\be
\ba{l}
\tau^{(2)} p_k -p_k \tau^{(2)}=0, \qquad \tau^{(2)} {\mathcal H}+{\mathcal H} \tau^{(2)}=0,
\vspace{1mm}\\
\tau^{(2)} J_{kl}-J_{kl} \tau^{(2)}=0, \qquad
\tau^{(2)} J_{0k}(-t_0)+J_{0k}(t_0) \tau^{(2)}=0,
\ea
\ee
\be
\ba{l}
\tau^{(3)} p_k^* +p_k \tau^{(3)}=0, \qquad \tau^{(3)} {\mathcal H}+{\mathcal H} \tau^{(3)}=0,
\vspace{1mm}\\
\tau^{(3)} J^*_{kl}+J_{kl} \tau^{(3)}=0, \qquad
\tau^{(3)} J^*_{0k}+J_{0k} \tau^{(3)}=0,
\ea
\ee
\be
\ba{l}
[\theta, p_n]_-=0, \qquad \theta{\mathcal H}^*(-\overline p)+{\mathcal H}(\overline p)\theta=0,
\vspace{1mm}\\
\theta J_{nl}^*(-\overline x) +J_{nl}(\overline x)\theta =0,
\qquad
\theta J_{0n}^*(-\overline x) -J_{0n}(\overline x)\theta =0,\qquad
\theta\theta^*(-\overline p)=1,
\ea
\ee
\be
\ba{l}
[\theta^{(n)}, p_m]_\pm=0, \qquad \theta^{(n)}{\mathcal H}^*(-p_k)+{\mathcal H}(p_k)\theta^{(n)}=0,
\vspace{1mm}\\
\theta^{(n)} J_{ml}^*(- x_n) \pm J_{ml}(x_n)\theta^{(n)} =0,
\qquad
\theta^{(n)} J_{0m}^*(- x_n) \pm J_{0m}(x_n)\theta^{(n)} =0,
\vspace{1mm}\\
\theta^{(n)}(p_n)\left( \theta^{(n)}(-p_n)\right)^*=1,
\ea
\ee
where ``$-$'' is taken if $k=m$ or $k=l$.

With the help of definition (1.3.2)--(1.3.5), (D.1), (D.2) we find tee connection between the
operators $r^{(k)}$, $\tau^{(k)}$, $\theta$ defined on the sets $\{\chi\}$ and $\{\Phi\}$
\be
\ba{l}
\left\{ r^{(n)}\right\}^\Phi =U \left\{ r^{(n)}\right\}^\chi U^{-1}(\ldots,-p_n),
\vspace{2mm}\\
\left\{ r^{(n)}\right\}^\chi =U^{-1} \left\{ r^{(n)}\right\}^\Phi U(\ldots,-p_n);
\ea
\ee
\be
\left\{ \tau^{(1)}\right\}^\Phi =U \left\{ \tau^{(1)}\right\}^\chi {U^{-1}}^*,
\qquad
\left\{ \tau^{(1)}\right\}^\chi =U^{-1} \left\{ \tau^{(1)}\right\}^\Phi U^*;
\ee
\be
\left\{ \tau^{(2)}\right\}^\Phi =U \left\{ \tau^{(2)}\right\}^\chi U^{-1},
\qquad
\left\{ \tau^{(2)}\right\}^\chi =U^{-1} \left\{ \tau^{(2)}\right\}^\Phi U;
\ee
\be
\left\{ \tau^{(3)}\right\}^\Phi =U \left\{ \tau^{(3)}\right\}^\chi {U^{-1}}^*,
\qquad
\left\{ \tau^{(3)}\right\}^\chi =U^{-1} \left\{ \tau^{(3)}\right\}^\Phi U^*;
\ee
\be
\left\{ \theta\right\}^\Phi =U \left\{ \theta\right\}^\chi {U^{-1}}^*(-\overline p),
\qquad
\left\{ \theta \right\}^\chi =U^{-1} \left\{ \theta \right\}^\Phi U^*(-\overline p);
\ee
\be
\ba{l}
\left\{ \theta^{(n)}\right\}^\Phi =U \left\{ \theta^{(n)}\right\}^\chi {U^{-1}}^*(\ldots,-p_n),
\vspace{1mm}\\
\left\{ \theta^{(n)} \right\}^\chi =U^{-1} \left\{ \theta^{(n)} \right\}^\Phi U^*(\ldots, -p_n);
\ea
\ee
\be
\left\{ r\right\}^\Phi =U \left\{ r\right\}^\chi U^{-1} (-\overline p),
\qquad
\left\{ r \right\}^\chi =U^{-1} \left\{ r \right\}^\Phi U(-\overline p),
\ee
where, for example, $\{r^{(n)}\}^\Phi$ $\left(\{r^{(n)}\}^\chi\right)$ denotes the operator
$r^{(k)}$, defined on the set $\{\Phi\}$ $(\{\chi\})$.

From these relations it is seen that in a general case $r$, $r^{(k)}$, $\tau^{(k)}$, $\theta$,
$\theta^{(k)}$ are the operator functions dependent on $p_l$ and $\sigma_\mu$.

\medskip

\begin{enumerate}
\footnotesize

\item Fushchych W.I., {\it Nucl. Phys. B}, 1970, {\bf 21}, 321. \ \ {\tt quant-ph/0206077}

\item Shirkov Yu.M., {\it DAN USSR}, 1954, {\bf 694}, 857; 1954, {\bf 99}, 737.

\item Foldy L.L., {\it Phys. Rev.}, 1956, {\bf 102}, 568.

\item Shveber S., Introduction in relativistic
quantum field theory,  1963.

\item Shirkov Yu.M., {\it JETP (Sov. Phys.)}, 1957, {\bf 33}, 1196; \\
Lomot J.S., Moses H.E., {\it J. Math. Phys.}, 1962, {\bf 2}, 405.

\item Fronsdal C., {\it Phys. Rev.}, 1959, {\bf 113}, 1367.

\item Fushchych W.I.,, {\it Theor. Math. Phys.}, 1970, {\bf 4}, 360.

\end{enumerate}
\end{document}